\documentclass[12pt,preprint]{aastex}

\received{2002 August 30}
\begin{document}

\title{The topology of weak lensing field 
in the neighborhood of MS1054-03}

\author{Jun'ichi Sato{}$^{1}$, Keiichi Umetsu{}$^{2}$,
Toshifumi Futamase{}$^{3}$ \& Toru Yamada{}$^{4}$}
\affil{{}$^{1}$ Department of Aeronautics and Space Engineering,
Tohoku University,
Sendai 980-8579, Japan\\
{}$^{2}$ Institute of Astronomy and Astrophysics,
Academia Sinica, P.O. Box 23-141, Taipei 106, Taiwan, R.O.C\\
{}$^{3}$ Astronomical Institute, Tohoku University,
Sendai 980-8578, Japan\\
{}$^{4}$ National Astronomical Observatory of Japan, 
Mitaka 181-8588, Japan\\}

\begin{abstract}
We report the first measurement of genus curves 
for the two-dimensional mass map
in the neighborhood of rich, X-ray luminous galaxy cluster MS1054-03
at $z=0.83$,
reconstructed from weak lensing data obtained by Suprime-Cam
on the prime focus of 8.2m Subaru telescope
\footnote{ 
Based on 
data collected at Subaru Telescope, which is operated
by the National Astronomical Observatory of Japan. 
}.
We find that the genus curve measured in the whole survey field
deviates 
from that expected from a random Gaussian field.
We show that this non-Gaussianity is induced by the rich cluster
in this region, and that the genus curve for the region without the
cluster is consistent with the prediction for a random Gaussian field.
The results demonstrate the feasibility
of the genus statistics to examine the non-Gaussianity due to the large
scale structures and to
probe the statistical properties of the
large scale clustering.
\end{abstract}


\section{Introduction}\label{section1}
The origin and the evolution of the large scale structures attract much
attention 
and the understanding about the statistical nature of the
mass distributions in the universe
is one of the important issues to be addressed in 
the observational cosmology. 
The inflationary theory gives us a definite
prediction on the statistical properties, namely the random Gaussian  
nature of the initial density fluctuation. 
The density fluctuation evolves by gravitational instability 
to have the  non-linear distribution, i.e. non-Gaussianity, at later stage
 whose details may depend on the cosmological parameters as well as 
the nature of dark matter.
Thus the measurement of statistical properties 
of the density distribution  is very  important 
for the understanding of the formation and evolution of 
large scale structures. 

The topology of large scale structures was qualitatively described by
\citet{Einasto84} and its quantitative study was initiated 
by \citet{Gott86}, \citet{hamilton86}, and \citet{got87}, 
and has been studied by many authors on the basis of various 
observational data.
For example, \citet{col87}, \citet{mel89}, and \citet{got90} 
investigated the genus or the Euler-Poinca\'re characteristic 
statistics of the topology in 
two-dimensional fields of galaxy distribution.
Further, the method has been applied for 
various kinds of observational samples, 
the angular distributions  
of galaxies on the sky \citep{col91, got92}, 
the galaxy distributions in slices of the universe
\citep{PGMK92, col97} and 
the temperature fluctuation fields of the Cosmic Microwave Background
\citep{col88, got90, smo94, kog96, col96, par98}. 
Recently the topology of the galaxy distributions in 
2 degree Field Galaxy Redshift Survey and Sloan Digital Sky Survey 
has been studied by \citet{Hoyle02a} and \citet{Hoyle02b}. 
These studies show that the topology of the two-dimensional galaxy
distribution is consistent with the random phase 
Gaussian distribution with 
a possible weak 'meat-ball' topology where the topology is called 
 'meat-ball' when it looks like collection of isolated islands. 
More recently, \citet{park01} calculated the two-dimensional 
genus statistics for the contour maps of the galaxy distribution 
in the Hubble Deep Field. 
Using the red-shift distribution of galaxies 
measured by the method of photometric red-shift, 
they studied the evolution effect in the topology.
However, the genus curves was consistent with random Gaussian field, 
and they were not able to discover any meaningful red-shift dependence.

Previous topological studies 
using galaxy distributions suffer from 
the biasing effect between 'visible' galaxies and underlying mass
distributions and it is not clear whether 
the galaxy distribution traces mass distribution 
in the universe precisely.
Thus it would be very useful to develop a method to measure 
mass distributions in the universe directly.
We shall make use of weak gravitational lensing techniques
for this purpose. 
Weak lensing directly measures the projected mass distributions
from the observed image distortions of background galaxies and thus 
can be used to compare
the observational results directly with the theoretical
and/or numerical predictions. 
The reliability of the weak lensing techniques for
measuring the mass distributions on large scales 
has been widely recognized, for example, by cosmic shear measurements
\citep{Ludvic00a, Wittman, Bacon, 
KNL00, maoli01, rhodes01, rhodes02, Ludvic01a}.

Our method for addressing this problem
makes use of the genus curves for weak lensing maps of 
the convergence field.
\citet{STJF01} have studied quantitatively the method
based on two dimensional Minkowski functionals (the genus curve is one
of the Minkowski functionals) via $N$-body and ray-tracing simulations,
and demonstrated its feasibility in the determination of
the cosmological density parameter, $\Omega_m$.
Using  the two dimensional genus statistic,
\citet{MJ} also studied the topology of weak lensing fields, and
found that the genus curve for large smoothing angles agrees with 
the predictions from linear theory,
while that for smoothing angles smaller than  $10'$ shows 
the non-Gaussian signatures of 
gravitational clustering and differs for open 
and flat cold dark matter models.

In the present paper, we report the first measurement 
of genus curves for the two-dimensional mass map
in the neighborhood of distant rich galaxy cluster MS1054-03,
reconstructed from weak lensing data obtained by Suprime-Cam
(Subaru Prime Focus Camera) on $8.2$m Subaru telescope 
\citep{umetsu01,umetsu+02}.
We examine the non-Gaussian feature in the measured genus curve
by comparing it with a Gaussian prediction, and demonstrate the feasibility
of the method to examine the statistical properties of the 
large scale mass distributions in the universe.

\section{Observation}\label{section2}
\begin{figure}[ht]
\vspace{10cm}
\caption{
The weak lensing convergence map reconstructed from the 
image distortions of background galaxies in the $R_{\rm c}$-band
image of Suprime-Cam
with Gaussian smoothing 
scale of $0\farcm 39$.
The side length is $19\arcmin$ and pixel size is $0\farcm 074$. 
The four sub-regions are also indicated in the figure.
The rich galaxy cluster is contained in the region 3, 
indicated by cross.
}
\includegraphics{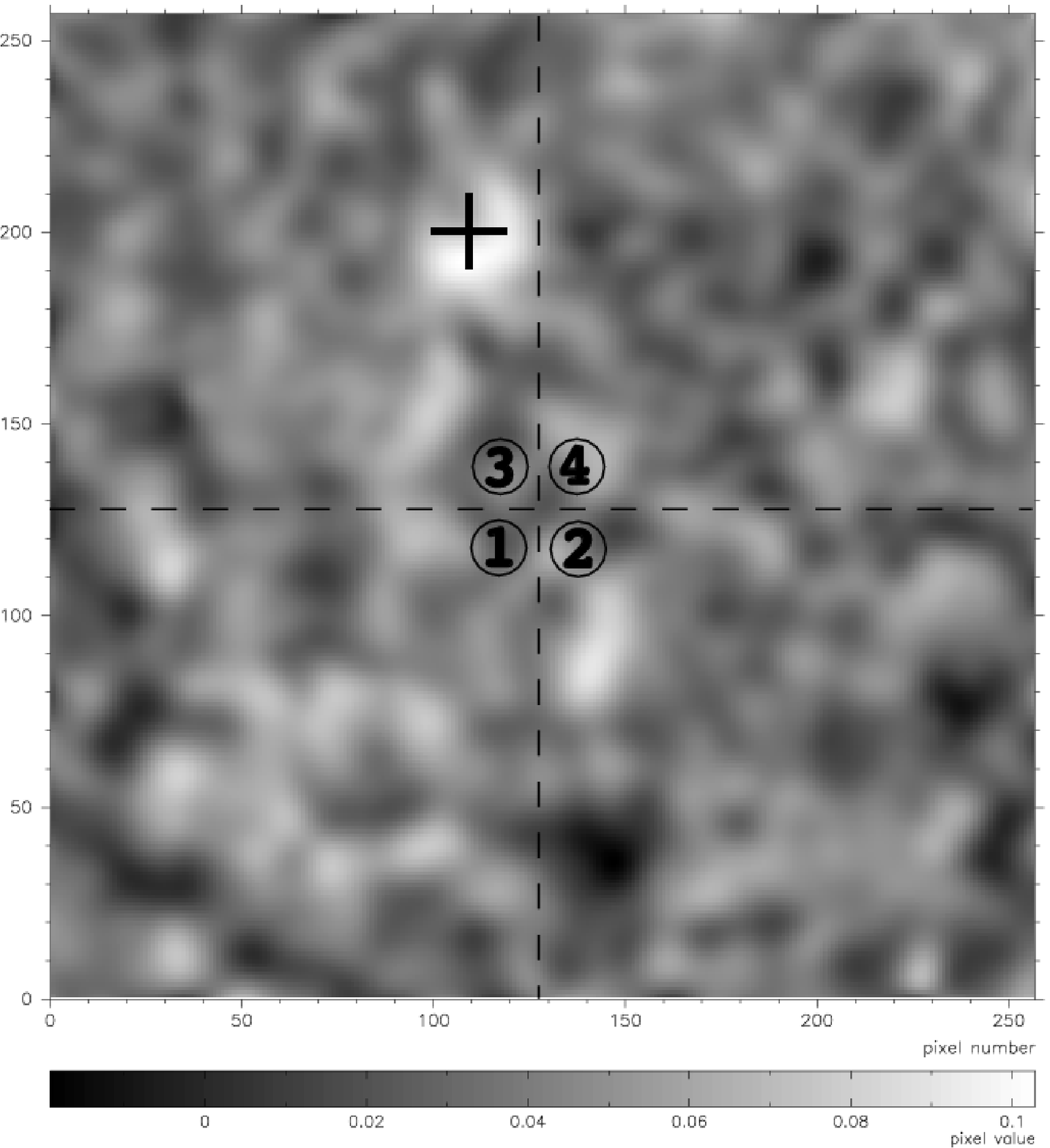}
\label{fig_1}
\end{figure}

We observed the rich, X-ray-luminous galaxy cluster
MS1054-03 at $z=0.83$ on the night of 2001 March 29 
using Suprime-Cam mounted at the prime focus of $8.2$m Subaru
telescope \citep{umetsu01,umetsu+02}.
We obtained a mosaic of $R_{\rm c}$-band images 
with a total exposure time of
$4.6$ ksec, which covers the sky area of 
$\sim 30 \times 30$ arcmin$^{2}$
with the seeing FWHM of $0\farcs 8$.
For weak lensing analysis, we selected galaxies of $R_{\rm c}<26$ with
signal-to-noise ratio (S/N) greater than
$6$ in the central $20\times 20$ arcmin$^2$ region.
The number of objects in the weak-lensing 
catalog is $11609$ ($29.2$ arcmin$^2$).
We have reconstructed the projected mass distribution
from the image distortions of sample galaxies
via Kaiser \& Squires method \citep{KS93, utf99} in Fourier space.
We first calculated the Gaussian smoothed shear field with 
$\theta_{\rm_{G}}=0\farcm 39$ 
from the weak lensing catalog, 
where the Gaussian window is defined by 
$W_{\rm{G}}(\theta)=\exp(-|\vec{\theta}|^{2}/2\theta_{\rm{G}}^{2})
/2\pi \theta_{\rm{G}}$.
The shear map is then 100\% zero-padded on each side
to avoid spurious effects from periodic boundary condition.
In Figure \ref{fig_1}, we show the reconstructed convergence map with 
a side length of $19\arcmin$ and pixel size of $0\farcm 074$.
The reconstructed convergence map shows a significant 
mass peak with S/N of $5.4$
associated with the rich cluster MS1054-03.


\section{Genus Statistic}\label{section3}
The genus statistic provides a topological description of large scale 
structures in the universe.
Since the weak lensing field is mapped on the sky, 
the two dimensional genus statistic is used in this case. 
For the local convergence field $\kappa$ shown in Figure \ref{fig_1},
we consider contour lines of a threshold 
value labelled by $\nu$. The threshold $\nu$ is usually 
taken as $\kappa = \nu \sigma_{0}$ for random Gaussian field, 
where $\sigma_{0}$ is the rms of the field, 
$\sigma_{0}^{2} = \left<\kappa^{2}\right>$.
The genus curve is expressed as function of 
this threshold level $\nu$.
The two dimensional genus $G_{2}$ is intuitively defined as 
\citep{adl81, col88, mel89, got90},
\begin{eqnarray}
&& G_{2}=(\mbox{number of contours surrounding regions}
\nonumber \\
&& \qquad \qquad \qquad \qquad
\mbox{higher than the threshold value})
\nonumber \\
&& \qquad -(\mbox{number of contours surrounding regions}
\nonumber \\
&& \qquad \qquad \qquad \qquad
\mbox{lower than the threshold value}).
\label{eq_1}
\end{eqnarray}

For a random Gaussian field, the genus curve is 
analytically given by \citep{matsubara00},
\begin{equation}
G_{2}(\nu)=
\frac{1}{2(2\pi)^{3/2}}\left(\frac{\sigma_{1}}{\sigma_{0}}\right)^{2}
\nu e^{-\nu^{2}/2},
\label{eq_2}
\end{equation}
where  $\sigma_{1}$ is defined by 
$\sigma_{1}^{2} \equiv -\left<\kappa \nabla^{2}\kappa \right>$.

Let us consider the meaning of value of the genus. 
If the value of genus is positive, then there is more 
isolated clusters than isolated voids and thus 
we can get an image that the topology 
is collection of isolated islands. 
So, when the value of genus takes large positive value, 
the topology is compared to 'meat-ball'.
On the other hand, if the value of genus is negative, 
there is less isolated clusters than isolated voids.
In this case the topology gives an image 
of the complicated shape with a lot of holes, and the topology 
is compared to 'Swiss cheese'. Thus we say 
the genus curve shows a  'meat-ball shift' and 'Swiss cheese shift' 
when the curve deviates from Gaussian shape to positive direction
and to negative direction, respectively.

Gaussian expression of 
equation \ref{eq_2} is not expected for the present 
weak lensing data because the survey area is 
not large enough to probe the linear scales.
However, it is known that the shape of the genus curve 
of mass distribution fields is not strongly affected by nonlinear 
gravitational evolution as shown by 
\citet{GWM87}, \citet{WGM87} and \citet{MWG88}. 
Thus we use the method of rescaled labeling of $\nu$.
For non-Gaussian fields, the threshold is defined through 
the fraction of area $f$ on the high-density side of 
the contour lines, which is related to the corresponding 
value of $\nu$ for a Gaussian field as, 
\begin{equation}
f=\frac{1}{\sqrt{2 \pi}}\int^{\infty}_{\nu}e^{-t^{2}/{2}}dt.
\label{eq_3}
\end{equation}
For a Gaussian field, this definition gives the same result 
as specifying the threshold directly in units of $\sigma_{0}$.
On the other hand, for a non-Gaussian field, 
it compensates for the horizontal shift of the genus curve 
due to the shift of  distribution function of the density field.
We adopt this prescription for setting the threshold values 
of $\kappa$ in the results shown below.

To calculate the genus curves for the convergence
map given as pixel data, 
we employed the method developed by 
\citet{Winit}, \citet{NFS99} and \citet{Schmalzing99}.
We confirmed that this code gives the same result as 
from CONTOUR2D developed by \citet{melott89}.
At each threshold level $\nu$,
the genus is obtained as an average over three genus values
with $\nu$s shifted by 0 and $\pm$0.1 as done in \citet{park01}.


\section{Results}\label{section4}
\begin{figure}[ht]
\vspace{10cm}
\caption{
Genus curve per steradian of the convergence field 
shown in Figure  \ref{fig_1}.
The open squares and dashed-curve indicate the genus 
measurements from the observational data.
The solid curve indicates the random Gaussian prediction
with best-fit normalization to the measured genus curve.
}
\includegraphics{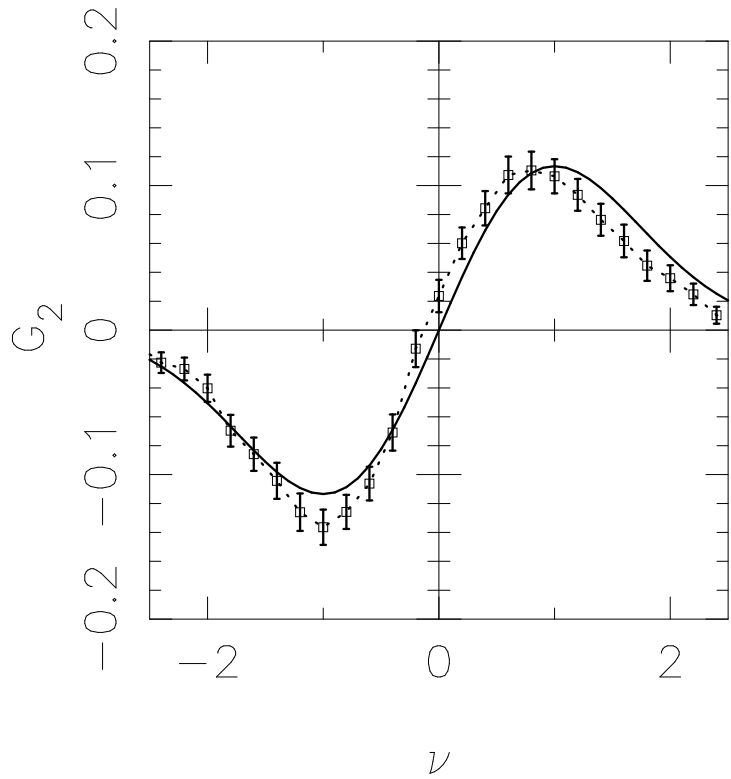}
\label{fig_2}
\end{figure}

Figure \ref{fig_2} shows the genus curve measured from 
the convergence field shown in Figure \ref{fig_1}.
The Gaussian prediction curve  is also calculated
by equation \ref{eq_2}, with the normalization factor obtained from
the least-$\chi^2$ fitting
to the measured genus curve.
The errors for the genus measurements are estimated from 
$100$-bootstrap resamplings of the background galaxies. 
We evaluated the curvature only at interior vertices 
for genus calculation.
Figure \ref{fig_2} shows a clear difference
between the random Gaussian prediction curve 
and the curve measured from the observational data. 
Furthermore, we found that compared with the random Gaussian 
prediction, the observed curve was shifted upwards 
near the threshold of $\nu=0$.
This indicates that the number of isolated clusters dominates over 
number of isolated voids, which is the so called 'meat-ball shift'.
This phenomenon is usually seen 
when the field evolves into non-Gaussian from Gaussian. 
However, it is difficult to judge 
from this result shown in Figure \ref{fig_2} 
whether this non-Gaussianity for the genus curve is 
influenced by the rich cluster contained in this region or 
by the large scale structures.

\begin{figure}[ht]
\vspace{10cm}
\caption{
Same as Figure \ref{fig_2} but
for each of the four sub-regions indicated in Figure \ref{fig_1}.
}
\includegraphics{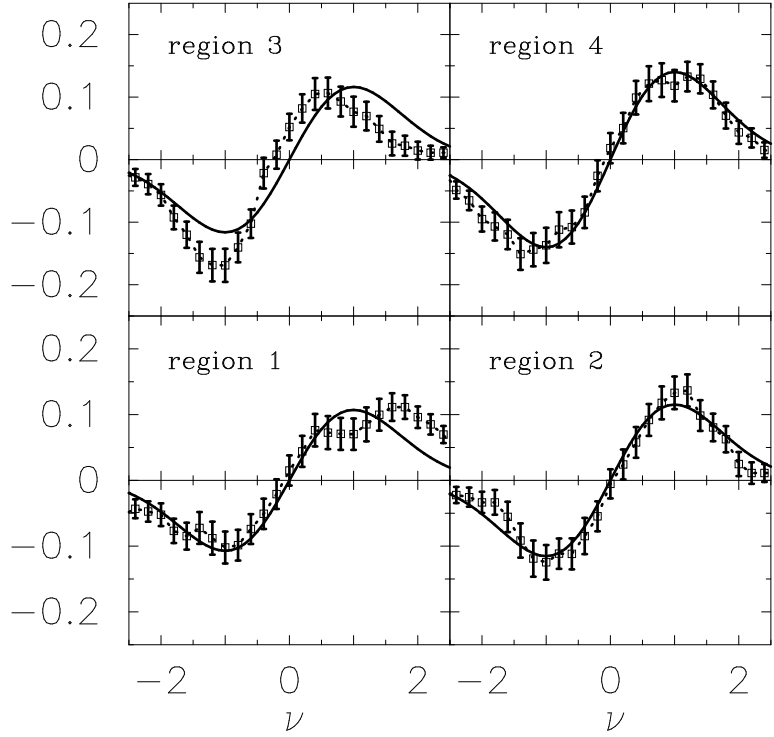}
\label{fig_3}
\end{figure}

To clarify the origin of the non-Gaussian nature, 
we divided the reconstructed convergence map into four square regions
with the same size, 
and calculated the genus curve for each sub-region.
As indicated in Figure \ref{fig_3}, we call each sub-region as 
region 1, 2, 3, and 4,
respectively. The rich cluster MS1054-03 is contained in region 3.

Figure \ref{fig_3} shows the genus curves 
for the sub-regions and their best-fit Gaussian genus curves,
obtained in the same way as for the whole field.
From Figure \ref{fig_3}, 
we see that the genus curve for the region 3  
has a large skew and a 'meat-ball shift'. 
On the other hand, the genus curves for the regions 2 and 4 
are consistent with random Gaussian curves.
The genus curve for the region 1 shows a slight excess in
the high threshold region.
Our weak lensing analysis shows a significant mass concentration
associated with the rich cluster,
therefore the highly non-Gaussian feature in the genus curve
could be induced by this rich cluster in the region 3.

The skewness is also a powerful statistical tool
to measure the deviation from a random Gaussian distribution,
defined by
$\equiv \left<\kappa^{3}\right>/\left<\kappa^{2}\right>^{2}$.
We summarized in Table \ref{tab_1} the 
skewness measured for each sub-region.
We see the skewness for the region 3 is quite larger than
that for any other region.

\section{Discussion and Conclusions}\label{section5}

\begin{figure}[ht]
\vspace{10cm}
\caption{
Same as Figure \ref{fig_2} but
for the region without the rich galaxy cluster MS1054-03, 
obtained by averaging the genus curves for the regions 1, 2, and 4
by inverse variance weighting at each threshold bin.
The genus curve is consistent with the random Gaussian prediction.
}
\includegraphics{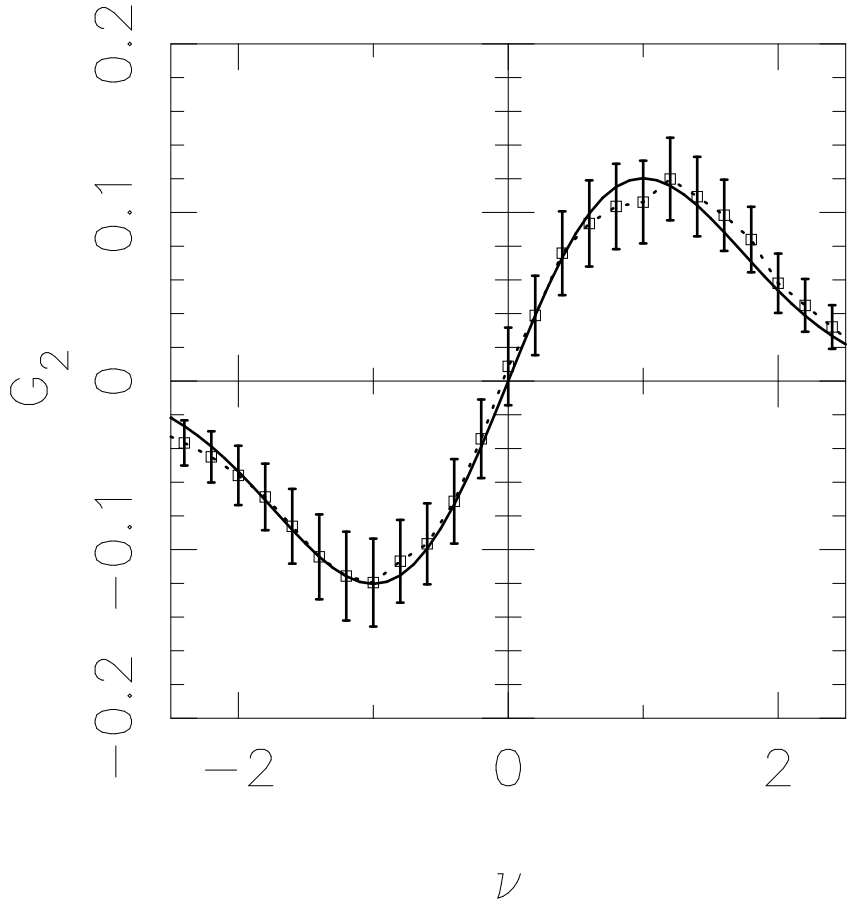}
\label{fig_4}
\end{figure}

We have measured for the first time the genus curves of the
convergence map reconstructed from image distortions of 
faint background
galaxies, as the first step for addressing the statistical properties
of the mass distributions on large scales.
The genus curve measured in the whole survey field (Figure \ref{fig_1})
shows a deviation from the random Gaussian prediction, 
i.e. non-Gaussianity of the field. 
By dividing the whole map into four independent
regions and analyzing them
separately,
we showed that the non-Gaussianity is induced by 
the rich cluster contained in the survey region (region 3). 
On the other hand, as for the regions without the rich cluster,
we also found a slight excess of the genus amplitude
in the high-thresholds for region 1, whereas the genus curves 
for regions 2 and 4 are consistent with random Gaussian distributions.
We thus averaged the genus curves of the regions 1, 2, and 4 by inverse 
variance weighting at each threshold bin. 
The resulting genus curve shown in Figure \ref{fig_4}
is consistent with the random Gaussian one at each threshold bin
within $1 \sigma$ error. We thus conclude the excess of the genus 
curve in region 1 is not due to the non-Gaussianity but is caused 
by the sample variance due to the 
small area of the region as discussed by \citet{cgpark01}.

Our results might be influenced 
by the choice of smoothing angle ($\theta_{\rm{G}}=0\farcm39$ in the paper). 
The effective number of galaxies used for the local shear average
decreases as smoothing angle decreases,
and then the reconstructed convergence map is much influenced by the
random noise due to the intrinsic distribution of galaxy ellipticities.
Thus the genus curves approach to 
the random Gaussian predictions as smoothing angles become smaller. 
\citet{MJ} discussed the effects of the smoothing angle 
on the genus measurements by 
the simulations with survey area of about $10$ degree$^{2}$ .
They concluded that for the smoothing angle of $\theta_{\rm G}=1\arcmin$, 
the genus curve shows a non-Gaussian 'meat-ball shift'.
On smoothing angles smaller than $1\arcmin$, 
it is not clear yet whether the observed non-Gaussianity is real or not.
We have not used 
$\theta_{\rm G}=1\arcmin$ because 
it could decrease the number of independent samplings,
and thus affect our analysis. 

In order to clearly see the deviation due to large scale structure, 
we definitely need to observe much wider region which will be planned 
in the near future survey by Suprime-Cam on Subaru telescope. 
The present study demonstrates the feasibility 
of the genus statistic for the weak lensing mass maps
to examine directly the statistical properties of the 
large scale mass distributions in the universe.
\\[5pt]

\section*{Acknowledgements}
We are very grateful to
Masahiro Takada and Chan-Gyung Park for useful discussions.

\clearpage

\clearpage

\begin{table}
\caption[]
{Skewness of the convergence field 
measured for the whole field and 
the four sub-regions shown in Figure \ref{fig_1}.
The errors are estimated from $100$ bootstrap 
resamplings of galaxies. }
\centering
\begin{tabular}{c|*{2}{c}}
\hline
region & skewness \\
\hline
whole &  8.53 $\pm$ 5.99 \\
1 &  4.68 $\pm$  7.42\\
2 &  1.62 $\pm$ 11.32\\
3 & 19.45 $\pm$ 10.02\\
4 &  2.42 $\pm$ 10.76\\
\hline
\end{tabular}
\label{tab_1}
\end{table}

\end{document}